# Influence of shear waves on transcranial ultrasound propagation in cortical brain regions

Ya Gao[1,2,3], Beat Werner[4], Beatrice Lauber[5], Yiming Chen[1,2,3], Giovanni Colacicco[5], Daniel Razansky[1,2*], Héctor Estrada[1,2*]

[1] Institute for Biomedical Engineering, Department of Information Technology and Electrical Engineering, ETH Zurich, Zurich, Switzerland

[2] Institute of Pharmacology and Toxicology, Faculty of Medicine, University of Zurich, Zurich, Switzerland

[3] Institute of Acoustics, School of Physics Science and Engineering, Tongji University, Shanghai, China

[4] Center for Magnetic Resonance (MR) Research, University Children's Hospital Zurich, Zurich, Switzerland

[5] Institute of Anatomy, Faculty of Medicine, University of Zurich, Zurich, Switzerland

* Correspondence: D. Razansky (daniel.razansky@uzh.ch) or H. Estrada (hector.estrada@pharma.uzh.ch)

*Abstract*— Transcranial ultrasound applications require accurate simulations to predict intracranial acoustic pressure fields. The current gold standard typically consists of calculating a longitudinal ultrasound wave propagation using a fluid skull model, which is based on full head CT images for retrieving the skull's geometry and elastic constants. Although this approach has extensively been validated for deep brain targets and routinely used in transcranial ultrasound ablation procedures, its accuracy in shallow cortical regions remains unexplored. In this study, we explore the shear wave effects associated with transcranial focused ultrasound propagation, both numerically and experimentally. The intracranial acoustic pressure was measured at different incidence angles at the parietal and frontal regions in an *ex vivo* human skull. The fluid-like skull model was then compared to the solid model comprising both longitudinal and shear waves. The results consistently show a larger error and variability for both models when considering an oblique incidence, reaching a maximum of 125% mean deviation of the focal area when employing the fluid skull model. Statistical assessments further revealed that ignoring shear waves results in an average ~40% overestimation of the intracranial acoustic pressure and inability to obtain an accurate intracranial acoustic pressure distribution. Moreover, the solid model has a more stable performance, even when small variations in the skull-transducer relative position are introduced. Our results could contribute to the refinement of the transcranial ultrasound propagation modeling methods thus help improving the safety and outcome of transcranial ultrasound therapy in the cortical brain areas.

## I. Introduction

Transcranial ultrasound (tUS) provides a technology platform for various brain applications. It has been proven that low-intensity ultrasound can produce neuromodulation effects in different parts of the brain [1]–[3]. When combined with microbubbles, tUS can disrupt the blood-brain barrier in a local, safe, and reversible manner to achieve local drug delivery [4]–[6]. The thermal and mechanical effects of high-intensity focused ultrasound (HIFU) are utilized for non-invasive clinical treatments of functional brain disorders, such as essential tremor [7] and Parkinson disease [8], [9]. Yet, accurate delivery of the desired ultrasound power through the intact skull remains a major challenge. Significant attenuation and distortion arise not only from the acoustic impedance differences between the skull and the brain, but also due to the skull's heterogeneity, multiscale porosity, and variability within and among individuals [10], [11]. Hence, case-specific simulations are needed to perform tUS interventions safely.

The accuracy of intracranial acoustic field predictions is directly dependent on the skull – transducer geometry and the elastic properties of the skull. Most simulation methods [12] neglect shear waves and thus simulate fluids skulls. This approximation is particularly convenient in terms of computational cost and speed. Despite the success of the fluid skull approach when the focus is located close to the center of the cranial vault [7]–[9], the role of shear waves in tUS has been found relevant at high incidence angles [13] or closer to the skull [14]. Subsequent studies have shown that oblique angle incidence could improve transmission efficiency and that shear waves are not negligible [13], [15]–[19]. Pichardo et al. predicted intracranial acoustic pressure based on the 3D viscoelastic wave simulations validated with ex vivo cranial experiments [20], although no transcranial focusing was attempted.

The effect of shear waves for shallow focusing at normal incidence has been numerically investigated using 2D simulations [21]. In addition, errors induced by the acoustic properties of the skull were explored in a shallow focus scenario and the effect of shear waves was investigated in 2D simulations at normal incidence [22]. However, experiments in a real skull were not reported. Yet, while shallow focusing is crucial for ultrasonic neuromodulation, the effect of shear waves in this context has not been verified experimentally. Thus, the error caused by neglecting shear waves at shallow focusing with different incidence angles remains a matter of uncertainty and warrants further research.

In this study, we explore the effect of shear waves in transcranial transmission and experimentally verified the accuracy of three-dimensional simulations at different incidence angles. In addition, we analyzed the spatial variation of the skull's insertion loss to provide more insight on the limits of current tUS measurement and simulation methods.

## II. METHODS

### A. Skull Preparation

The experiments were performed on a human skull *ex-vivo* following the regulations of the Swiss Association of Research Ethics Committees and approved by the Kantonale Ethikkommission Zürich (BASEC-Nr 2022-01238). The skull (87 years old, male) was preserved in Thiel's solution [23], [24] and degassed in saline solution in a vacuum chamber for at least 24 hours before any measurements. X ray-CT scans of the skull were acquired using Revolution CT / GE MEDICAL SYSTEMS at 120 kVp and reconstructed using bone window filtering with 391 × 391 × 625 µm$^3$ resolution and 512 × 512 in-plane matrix.

### B. Pressure Mapping Experiments

We measured the transcranial acoustic pressure fields at normal incidence and 40°-angle incidence **(Fig. 1)**. Measurements were done by a focused transducer (Imasonic A101, 0.55 MHz center frequency, 80 mm diameter, and 64 mm radius of curvature) and a 1 mm diameter polyvinyl difluoride (PVdF) needle hydrophone (Precision Acoustics, UK). The transducer is connected to a signal generator with a power amplifier (ENI 2100 L RF) and emits a pulse signal with a center frequency of 500 kHz, and a repetition frequency of 10 kHz. The hydrophone was mounted on a three-axis computer-controlled positioning system and scanned a 35 × 35 mm² scanning area in the orthogonal XY and XZ planes via an 'S' path. In the X-axis we use uninterrupted scanning and real-time acquisition, and the Y-axis and Z-axis are in steps of 0.3 mm.

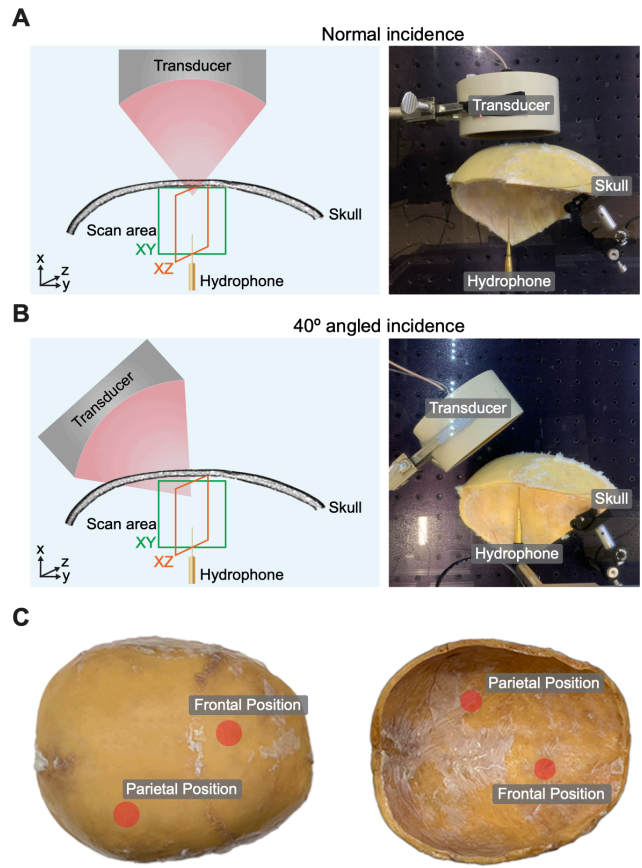

**Fig. 1.** Experimental setup at normal (A) and 40°-angle incidence (B). The photographs taken during the measurements are shown on the right. (C). The measured positions in the skull sample.

We measured two positions located in the parietal bone and frontal bone regions (**Fig. 1C**) and placed the transducer in the normal and 40° incidence positions, respectively. The thicknesses at the measurement positions of the parietal and frontal bones were around 5.2 mm and 6.1 mm. After each measurement, the skull was removed, and the ultrasound field in the free field was acquired in the same conditions. All measurements were done in a tank filled with degassed and deionized water.

*C. Skull Acoustic Properties*

The skull's geometry was extracted from the CT image by thresholding from 0 to 2000 HU. The distributions of acoustic properties of CT images (**Fig. 2A**) were obtained based on the relationship between CT Hounsfield values and density $\rho$, longitudinal $c_L$, and shear wave sound velocity $c_S$ [25]. The shear wave speed was fitted based on the relationship with density within the range of skull shear wave velocities between 1300 to 1540 m/s [13], and the shear wave is only set at a density greater than 1200 kg/m³ and is set to a constant value of 1540 m/s at a density greater than 2000 kg/m³. The density and longitudinal sound velocity of the surrounding soft tissue were set to 1000 kg/m³ and 1497 m/s. The acoustic properties mapped from CT image followed these equations [25]:

$$\rho = 1000 + 1.2HU$$

$$c_L = 167 + 1.33\rho$$

$$c_S = 967 + 0.27\rho$$

*D. Acoustic Simulations*

We performed the acoustic simulations using the k-Wave toolbox [26]. The simulation setup (**Fig. 2B, C**) consists of a focused transducer targeted transcranially at cortical brain regions. The acoustic simulation domain was 320 × 256 × 384 grid points with grid point spacing of 391 μm and extending for approximately 125×100×150 mm³. The skull's geometry was obtained by segmenting from the CT image and cropping the surrounding excess background area, then resliced and placed into the 3D simulation grid. The transducer used in the experiments was modelled using a 0.5 MHz focused transducer (~36% bandwidth) with 64 mm curvature radius and 80 mm diameter. The simulation ended at 100 μs with a 0.3 Courant-Friedrichs-Lewy (CFL) number. All computations were performed on GPU (NVIDIA Corporation GA102 [GeForce RTX 3090]), with a single 3D elastic wave simulation taking about 15 minutes.

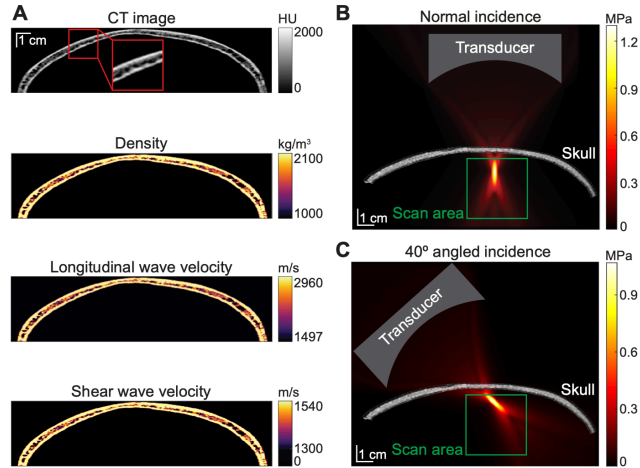

**Fig. 2.** (A) The acoustic properties (density $\rho$, sound speed of longitudinal wave $c_L$, and sound speed of shear wave, $c_S$) map of the skull were calculated using HU values in the CT images. Simulation setting in k-Wave toolbox at normal incidence (B) and 40° angled incidence (C) with the pressure map in free field. The focus targets shallow cortical brain regions.

*E. Data Processing and Simulation Registration*

Sliding average processing on the X-axis and frequency filtering below 0.1 MHz and above 1 MHz by Zero-phase digital IIR filtering were performed on the acquired experiment data. We then reconstructed the acoustic field in scan area of the XY and XZ planes with a resolution of 0.3 mm × 0.3 mm × 0.1 μs.

For each acoustic field measurement, a corresponding simulation was performed. Regarding the registration of the simulations and experiments, based on our previously mentioned free-field measurements, we aligned the diameter, radius of curvature, emission angle of the transducer, transmission signal as well as the scanning area. The registration results of the free-field are shown in the supplementary (**Fig. S1**). In the experiments, we marked the measured position of the skull in both the outer and inner layers, and subsequently tagged the location of the point in the CT image and embedded it in the simulation grid. Then the skull geometric positions and properties were aligned according to the acoustic field with skulls.

With all conditions constant, we calculated the acoustic pressure field of the solid skull and correspondingly calculated the acoustic pressure field of the fluid skull by $c_S = 0$ [27]. The maximum acoustic pressure and the -6 dB area were used as evaluation metrics. In addition, since retracing the experimental marker points in the simulation may introduce position errors, we performed the same simulation calculations for the fluid and solid skull by randomly selecting 12 points in a 5 mm × 5 mm area near the marker point as the center and statistically counted the maximum and -6 dB area of the intracranial acoustic field.

## III. Results

### A. Parietal Bone

The intracranial sound pressure fields from the parietal bone in the two perpendicular planes XY and XZ of the transducer at normal incidence (**Fig. 3A, B**) show a very good agreement between measurements and simulations. At 40° incidence (**Fig. 3C, D**), the differences are more noticeable, particularly at the maximum pressure distribution at the XZ plane. Even at normal incidence the tUS beam has a slight directional deflection in the XY and XZ planes due to the three-dimensional shape of the skull (**Fig. 3A, B**). Under the same simulation conditions ignoring shear waves, the intracranial pressure and the area of the -6 dB region increased, which was seen in both planes of normal and angled incidence. The propagation of the short wave through perpendicular planes demonstrate the complexity of the phenomenon and the need of using full three-dimensional approaches (see supplementary videos).

Due to the skull's geometry, at 40° the XZ plane scan couldn't capture the focus (**Fig. 3D**). Therefore, the slight differences observed at the XY plane (**Fig. 3C**) create noticeable differences in the predicted pressure distribution using solid and fluid skull.

### B. Frontal Bone

Due to the increased thickness of the frontal bone (~6.1 mm) compared to the parietal bone (~5.2 mm) and a more pronounced curvature, the tUS wave suffers larger transmission losses through the frontal bone. As the focal region covers larger areas of the skull, the US wave is more affected by the curvature and thickness. The predicted peak pressure and -6 dB area are larger using the fluid skull model at normal incidence (**Fig. 4A, B**). At 40° incidence (**Fig. 4C, D**), the predicted peak pressure and the -6 dB area are close to the experiment in both simulations.

### C. Solid-fluid skull error statistics

Introducing a small variation (±5 mm) in the skull-transducer relative position allows us to compare the sensitivity of both solid and fluid skull models (**Fig. 5**). The intracranial pressure prediction for the solid skull consistently closer to the experiment at the parietal bone (**Fig. 5A, B**). In the fluid skull, the intracranial pressure is overestimated up to 40% (in the normal XZ plane) and reaches a maximum of about 100% (in the angled XY plane) at 40° incidence. In the frontal bone at normal incidence (**Fig. 5C, D**), the fluid skull intracranial pressure was 70% higher than the solid skull prediction (normal XZ plane). In general, the variability at 40° incidence is much greater than at normal incidence for both solid and fluid skull simulations, reaching up to 3.5 times the measured pressure (angled XY plane). The mean error in the relative maximum pressure and -6 dB area is over 40% at both normal and 40° incidence with the fluid skull (**Fig. 5E**). The standard deviation of the error under angled incidence is much larger than under normal incidence, but the fluid skull still has larger standard deviation under both incidence angles (**Fig. 5F**).

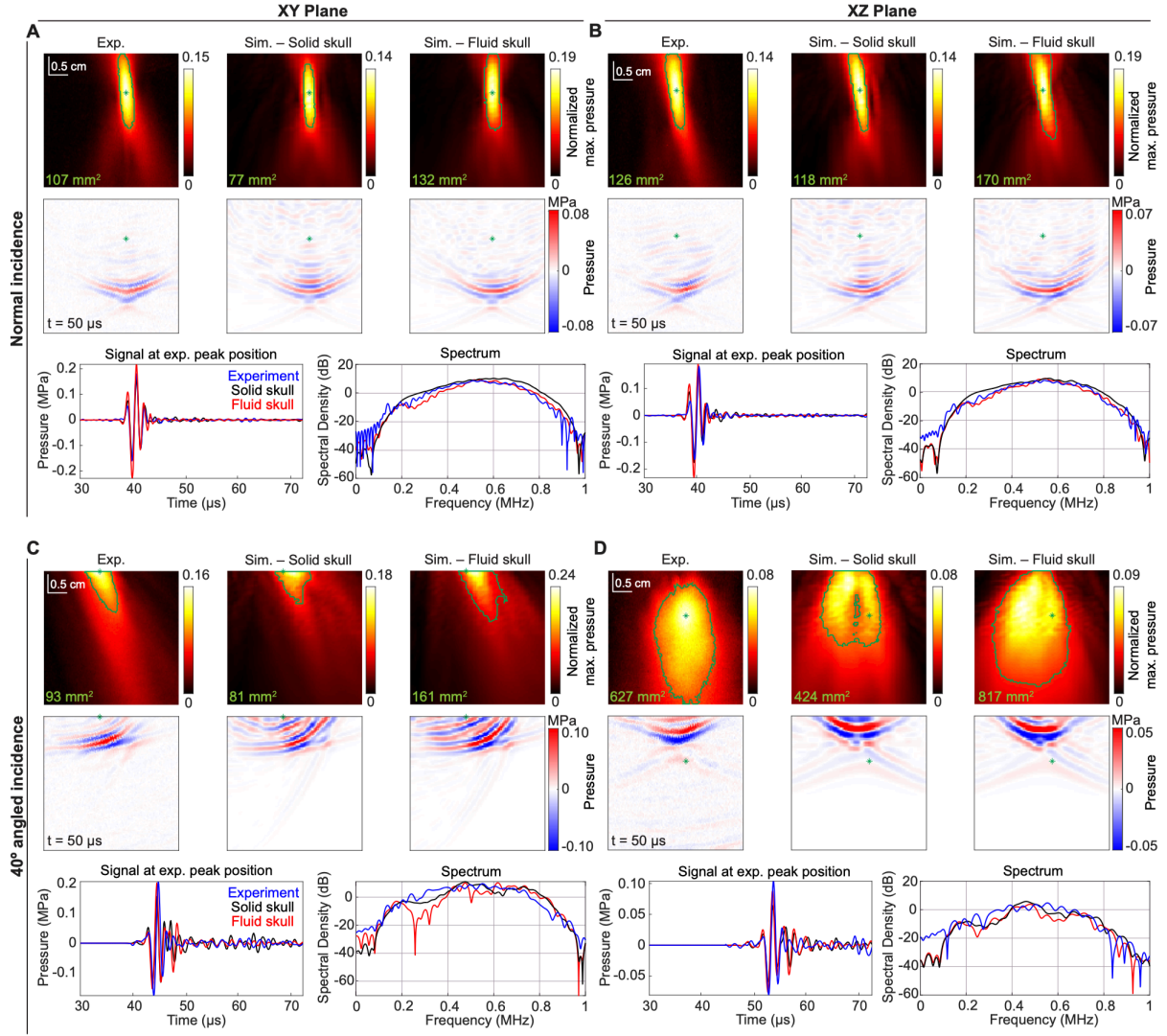

**Fig. 3.** Intracranial sound pressure fields through the parietal bone from measurements, solid skull (with shear wave) and fluid skull (without shear wave) simulations at normal incidence in the XY (A) and XZ (B) planes and at 40° incidence in the XY (C) and XZ plane (D). At each panel: (top) maximum pressure field normalized by the measured peak pressure in free field, (middle) snapshot of the US propagation at t = 50 µs, and (bottom) US signal and spectrum at the point of peak measured pressure. The -6 dB area is circled in green and its corresponding value in green at bottom left of each max. pressure map.

*D. Skull insertion loss measurements*

We calculated the skull insertion loss as $IL = 20\ log_{10}\left(\frac{P_0}{P_S}\right)$, where $P_0$ and $P_S$ are the Fourier transforms of the measured pressure without and with the skull, respectively. To mimic single point hydrophone measurements usually used to quantify the skull's attenuation [10], we found points F1 and F2 (**Fig. 6A, B**) at the respective pressure peaks without (**Fig. 6A**) and with (**Fig. 6B**) the skull. Choosing either point to estimate skull transmission properties generates different results (**Fig. 6C**) ranging from a 2.5 dB difference at 0.5 MHz up to 4.2 dB at 0.8 MHz. The skull insertion loss greatly fluctuates over the space due to beam shifts (**Fig. 6D**), particularly in regions close to the focus where local fluctuations encompass a range of ~40 dB difference in less than 5 mm. Using spatial averaging over the -6 dB area could help reduce such fluctuations (**Fig. 5C**, green curve).

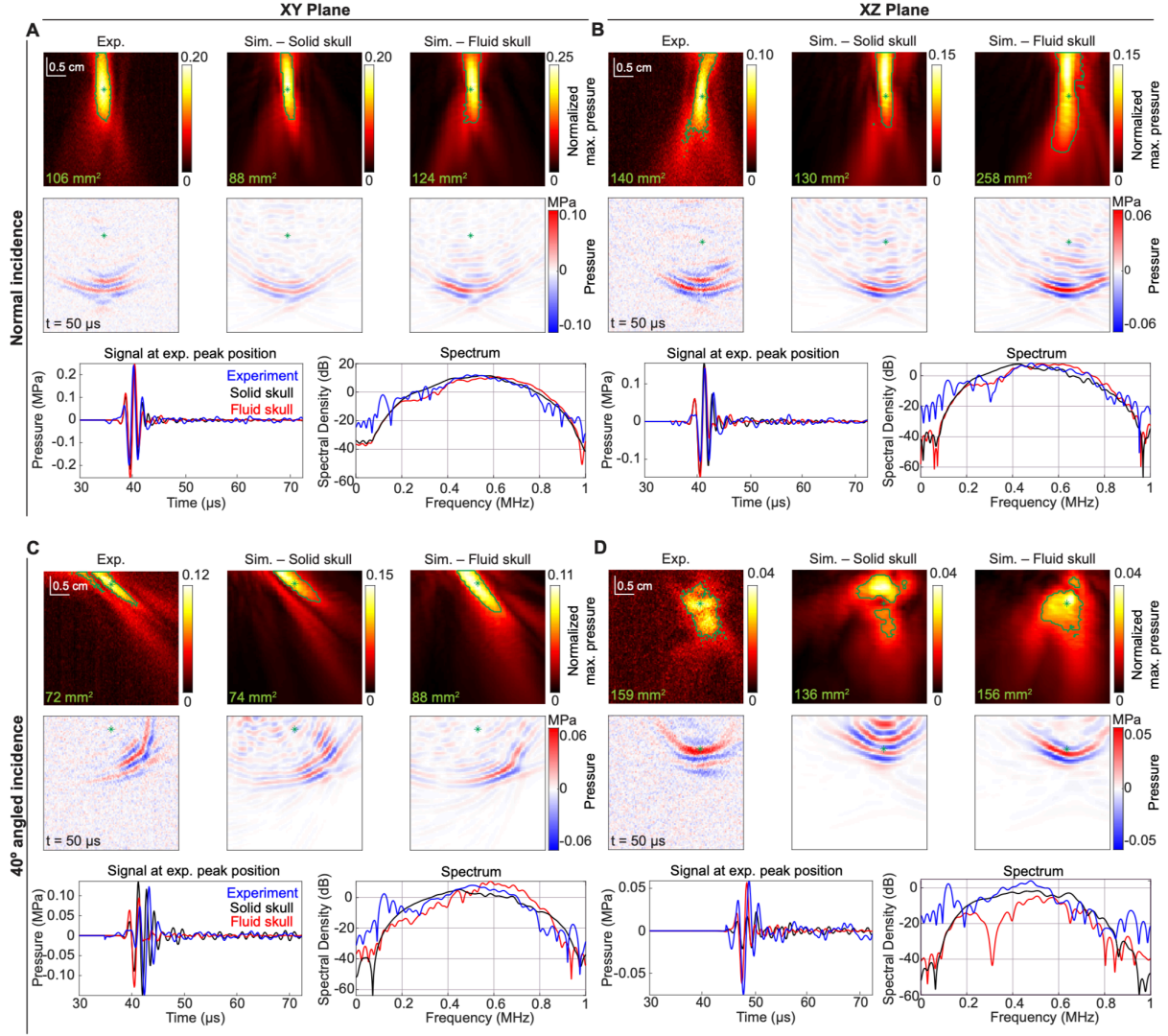

**Fig. 4.** Intracranial sound pressure fields through the frontal bone from measurements, solid skull (with shear wave) and fluid skull (without shear wave) simulations at normal incidence in the XY (A) and XZ (B) planes and at 40° incidence in the XY (C) and XZ plane (D). At each panel: (top) maximum pressure field normalized by the measured peak pressure in free field, (middle) snapshot of the US propagation at t = 50 µs, and (bottom) US signal and spectrum at the point of peak measured pressure. The -6 dB area is circled in green and its corresponding value in green at bottom left of each max. pressure map.

IV. DISCUSSION AND CONCLUSION

We examined the effect of shear waves in tUS by X ray-CT based 3D simulation and compared it against experiments on an ex-vivo skull and shallow focusing. We performed a sensitivity analysis of the effect of shear waves in different skull regions at normal and angled incidence. We found a consistent overestimation of the intracranial sound pressure when shear waves are ignored in fluid skull simulation. The deviations can be interpreted as shear waves contributing to the intracranial field by mode conversion to longitudinal waves before its transmission out of the skull, particularly due to the shallow focusing and the angular coverage of the transducer (77° full angle). At 40° incidence, the intracranial acoustic field is highly variable, presumably due to the longer propagation distances through the skull, thus making the intracranial acoustic field more sensitive to even small position changes.

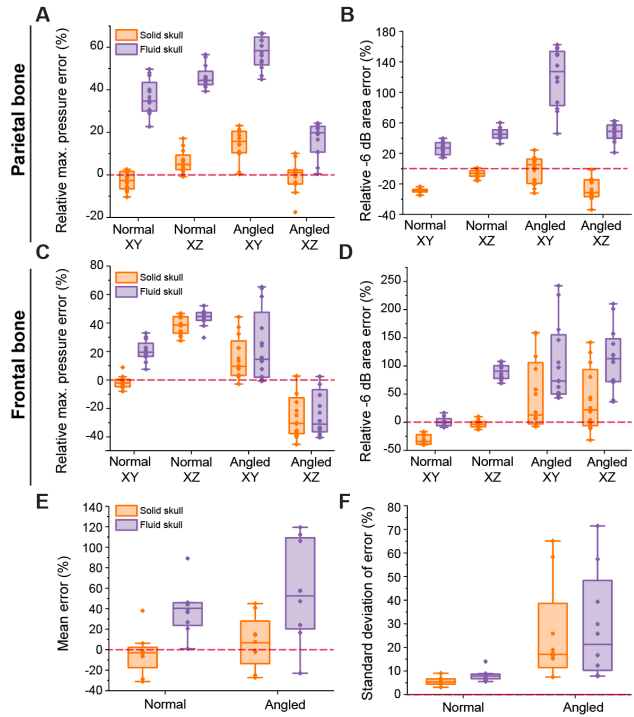

**Fig. 5.** Statistics of the predicted intracranial field error. Relative maximum pressure (A), (C), and -6 dB area (B), (D) error for both parietal and frontal bone. The mean (E) and standard deviation (F) of the errors at normal and angled incidence. For the box plots, the centre line represents the median, the box limits show the bottom and top quartiles, and the whiskers show the minimum to maximum values.

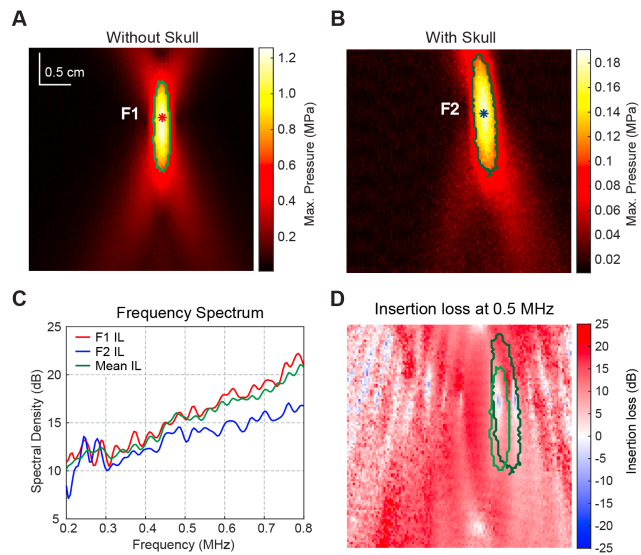

**Fig. 6.** Insertion loss of the parietal bone. Maximum pressure field without (A) and with (B) the skull. (C). Insertion loss as a function of the frequency. The red curve is calculated at the maximum pressure point F1 (red asterisk in (A)), while the blue curve at the maximum pressure point F2 (blue asterisk in (B)). The green curve is the average of the -6 dB area in (A) and (B). (D). Spatial distribution of insertion loss at 0.5 MHz.

Surprisingly, the inclusion of shear waves helped stabilize the intracranial acoustic field from positional changes even at normal incidence. Thus our results suggest that shear waves play a non-negligible role in the simulation of intracranial sound field for shallow focusing, generating a maximum of 90% error in normal incidence and 125% error in angled incidence.

In addition, it has been verified that shear waves are more affected by the fine structures within the skull due to the shorter wavelength of the shear wave compared to the longitudinal wave [28]. Simplifying by using homogeneous plates having the external shape of the skull only will introduce large structural errors. The use of CT-learned MR scans is also expected to increase the error [28]–[30].

In general, shear waves are ignored due to the significant increase in complexity and computation time of the numerical problem. The development of a highly accurate model that quickly predicts and takes into account shear waves is highly demanded by the research community and at the clinics. Deep learning is a powerful tool has been used for fast prediction of intracranial sound field in fluid skulls [11], [31], [32]. Therefore, it is very likely deep learning methods could help predict the intracranial sound field of solid skulls in the near future.

Skull characterization based on single point measurements is subject to large fluctuations. We found that scanning and using spatial averaging can help reduce such variations to provide a more robust characterization.

Our results are of immediate relevance for tUS applications requiring accurate localization and dosing at cortical regions. They highlight the importance of shear waves for the tUS simulations, as using the fluid skull assumption at normal and oblique incidence caused much larger errors than expected and had a non-negligible effect on both beam shape and pressure. In addition, our findings emphasize the importance of experiments in the future development of the tUS field. Despite the relatively large number of studies relying only on simulations [33], [34] and some solely on experimental characterization [35], bringing both together and performing measurement-based simulations [17], [20] would contribute to a better understanding of tUS propagation as well as enabling more accurate intracranial acoustic pressure prediction.

## V. Acknowledgment

The authors acknowledge support from the Personalized Health and Related Technologies (PHRT) grant of the ETH Domain (PHRT-582). This work was supported in part by the China Scholarship Council. We gratefully acknowledge D. Udovicic from the Dept. of Radiology, University Children's Hospital Zurich, for his technical support in acquiring CT data, and X. L. Deán-Ben for providing the power amplifier.

# Supplementary

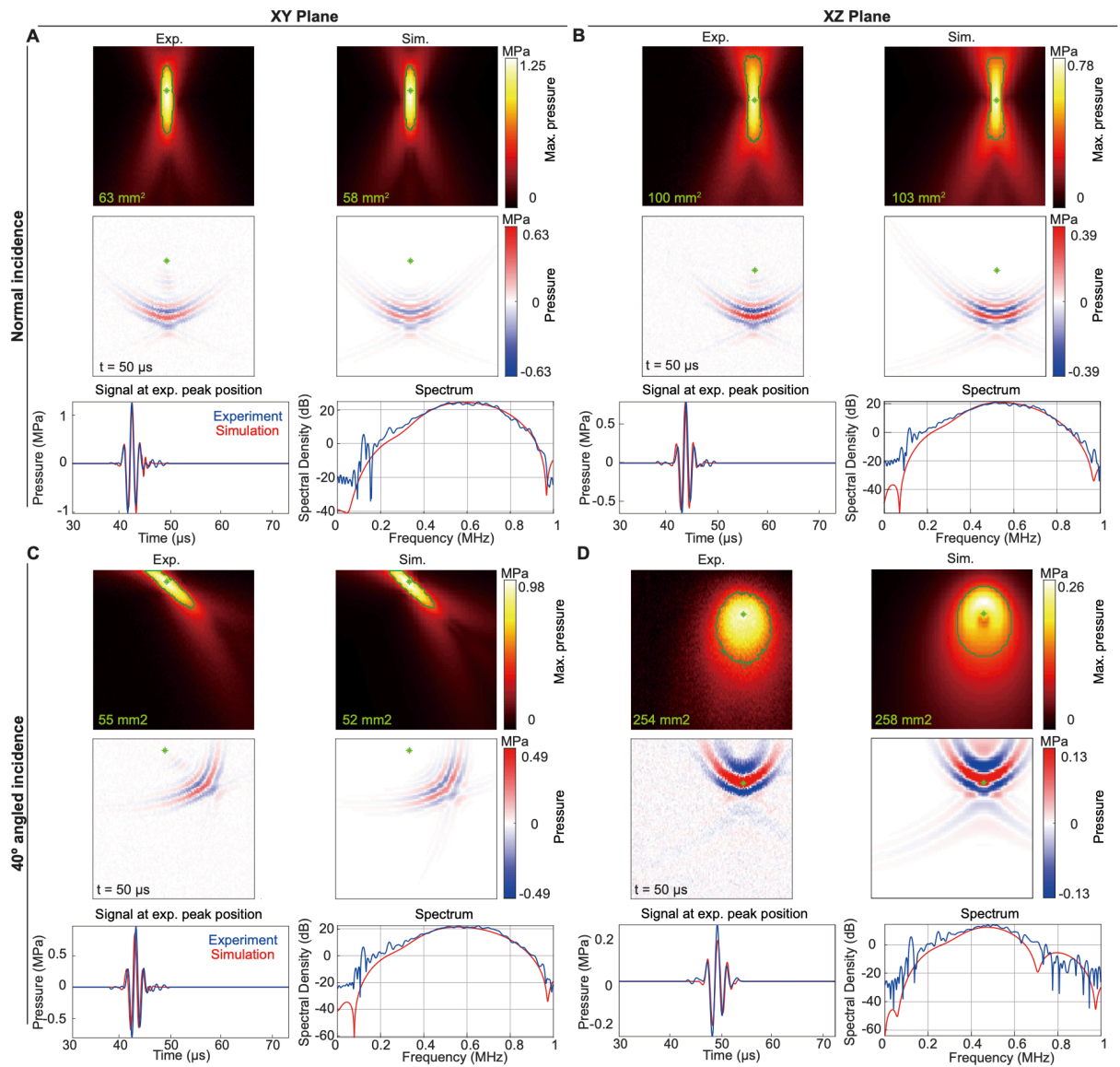

**Fig. S1.** Sound pressure fields without skull from measurements and simulations at normal incidence in the XY (A) and XZ (B) planes and at 40° incidence in the XY (C) and XZ plane (D). At each panel: (top) maximum pressure field normalized by the measured peak pressure in free field, (middle) snapshot of the US propagation at t = 50 µs, and (bottom) US signal and spectrum at the point of peak measured pressure. The -6 dB area is circled in green and its corresponding value in green at bottom left of each max. pressure map.